\documentclass[12pt]{article}
\usepackage{latexsym}
\usepackage{epsfig}
\begin{document}

\newcommand{\beq}{\begin{equation}}
\newcommand{\eeq}{\end{equation}}
\newcommand{\bear}{\begin{eqnarray}}
\newcommand{\eear}{\end{eqnarray}}
\newcommand{\Ka}{K\"ahler }
\newcommand{\half}{{{1}\over{2}}}
\newcommand{\e}{\epsilon}
\newcommand{\te}{\tilde{\epsilon}}
\newcommand{\htheta}{\hat{\theta}}
\newcommand{\nn}{\nonumber}

\newsavebox{\uuunit}
\sbox{\uuunit}
    {\setlength{\unitlength}{0.825em}
     \begin{picture}(0.6,0.7)
        \thinlines
        \put(0,0){\line(1,0){0.5}}
        \put(0.15,0){\line(0,1){0.7}}
        \put(0.35,0){\line(0,1){0.8}}
      
\multiput(0.3,0.8)(-0.04,-0.02){12}{\rule{0.5pt}{0.5pt}}
     \end {picture}}
\newcommand {\unity}{\mathord{\!\usebox{\uuunit}}}
\newcommand  {\Rbar} {{\mbox{\rm$\mbox{I}\!\mbox{R}$}}}
\newcommand  {\Hbar} {{\mbox{\rm$\mbox{I}\!\mbox{H}$}}}
\newcommand  {\Nbar} {{\mbox{\rm$\mbox{I}\!\mbox{N}$}}}
\newcommand {\Cbar}
    {\mathord{\setlength{\unitlength}{1em}
     \begin{picture}(0.6,0.7)(-0.1,0)
        \put(-0.1,0){\rm C}
        \thicklines
        \put(0.2,0.05){\line(0,1){0.55}}
     \end {picture}}}
\newsavebox{\zzzbar}
\sbox{\zzzbar}
  {\setlength{\unitlength}{0.9em}
  \begin{picture}(0.6,0.7)
  \thinlines
  \put(0,0){\line(1,0){0.6}}
  \put(0,0.75){\line(1,0){0.575}}
  \multiput(0,0)(0.0125,0.025){30}{\rule{0.3pt}{0.3pt}}
  \multiput(0.2,0)(0.0125,0.025){30}{\rule{0.3pt}{0.3pt}}
  \put(0,0.75){\line(0,-1){0.15}}
  \put(0.015,0.75){\line(0,-1){0.1}}
  \put(0.03,0.75){\line(0,-1){0.075}}
  \put(0.045,0.75){\line(0,-1){0.05}}
  \put(0.05,0.75){\line(0,-1){0.025}}
  \put(0.6,0){\line(0,1){0.15}}
  \put(0.585,0){\line(0,1){0.1}}
  \put(0.57,0){\line(0,1){0.075}}
  \put(0.555,0){\line(0,1){0.05}}
  \put(0.55,0){\line(0,1){0.025}}
  \end{picture}}
\newcommand{\Zbar}{\mathord{\!{\usebox{\zzzbar}}}}
\newcommand{\ew}{\ensuremath{\frac{1}{2}}}
\newcommand{\crond}{\ensuremath{\mathcal{C}}}
\newcommand{\krond}{\ensuremath{\mathcal{K}}}
\newcommand{\orond}{\ensuremath{\mathcal{O}}}

\begin{titlepage}
\begin{flushright} KUL-TF-98/29 \\ hep-th/9807095
\end{flushright}
\vskip 2.5cm
\begin{center}
{\Large \bf The rigid limit in Special \Ka geometry for $SU(2)$ SYM with a 
massive quark hypermultiplet} \\
\vskip 1.5cm
{\bf Chris Van Den Broeck$^\dagger$} \\
{\small Instituut voor Theoretische Fysica, \\
Katholieke Universiteit Leuven, B-3001 Leuven, Belgium }
\end{center}
\vskip 4cm
\begin{center}
{\bf Abstract}
\begin{quote}
We study the rigid limit of type IIB string theory, compactified on a $K3$
fibration, which, near its conifold limit, contains the Seiberg--Witten curve 
for $N=2$ $SU(2)$ Super-Yang-Mills with a massive hypermultiplet in the 
fundamental representation. Instead of working with an $ALE$ approximation, 
we treat the $K3$ fibration globally. The periods we get in this way, allow
for an embedding of the field theory into a supergravity model.
\end{quote}
\end{center}
\vfill
\hrule width 5cm
\vskip 2mm
{\small $^\dagger$ E-mail: chris.vandenbroeck@fys.kuleuven.ac.be}
\end{titlepage}

\section{Introduction}

In the past few years, a large number of quantum field
theories have been solved by the method of geometric engineering
\cite{geoengin}. Type IIB string theory compactified on a 
Calabi--Yau manifold, leads to an $N=2$ supersymmetric theory
in 4 dimensions. The vector multiplet moduli space of these 
theories recei ves no quantum corrections, and by going to a 
rigid limit of this classical moduli space and identifying the
corresponding rigid low energy quantum theory (usually a field 
theory), one is able to obtain an exact solution for 
the two derivative low energy effective action \cite{klemmreview,
lerchereview}. 

A type IIB compactification is mapped onto a type IIA 
compactification by mirror symmetry \cite{dixon, mirror}. This maps the 
non-quantum corrected vector moduli space of the type IIB model 
to the corresponding moduli space of the IIA model. The latter
does receive quantum corrections from world-sheet instantons.
When one needs to solve a quantum field theory, one can look for 
a type IIA model which yields this theory in a rigid limit,
and map it to a classical type IIB model by mirror symmetry.

When the Calabi--Yau has a conifold singularity, branes wrapped around
cycles which shrink to zero, become massless, which explains the
singularities in the moduli space of the effective quantum field theory
\cite{strominger}. Accordingly, the limit where the Calabi--Yau develops 
a conifold singularity is to be identified as a rigid limit. In principle, 
it is only necessary to study a neighbourhood of the singularity, and the 
Calabi--Yau can be approximated by an $ALE$ fibration \cite{klemm1, klemm2}. 
These fibrations make clear how the Seiberg--Witten curve of the
rigid model originates in string theory. However, this is not sufficient 
if we also want to embed the low-energy theory into a supergravity 
model. The supergravity theories we get in this way from Calabi--Yau 
manifolds which are $K3$ fibrations are important from a 
phenomenological point of view.

The two derivative action of the scalars in vector multiplets of
$N=2$, $D=4$ theories defines a geometric structure known as 
special \Ka geometry. There are two kinds: local special geometry 
\cite{dwlsvp,dwvp} applies to locally supersymmetric theories, 
i.e. supergravity and strings, while rigid special geometry \cite{sierra,
gates} is associated to rigid supersymmetry, i.e. $N=2$ supersymmetric 
gauge theories in flat spacetime. 

In \cite{rigidlim}, a detailed study was made of the way in which 
the rigid degrees of freedom decouple from the rest of the action 
in the rigid limit of the local geometry. The Calabi--Yau manifolds 
under investigation were $K3$ fibrations, which greatly facilitated 
the computations. The symplectic period vector of the Calabi--Yau, 
which is given in terms of integrals of the holomorphic 3-form over 
a basis of 3-cycles, could be written as a vector of $K3$ periods 
integrated over cycles in the base space of the fibration.

In the present paper, we exploit this split-up to study the rigid limit
of a Calabi--Yau that has the same $K3$ fibre as one of the models
studied in \cite{rigidlim}. It describes Seiberg--Witten theory
with a single massive quark hypermultiplet. 

The structure of this paper is as follows. In section 2, we 
introduce the polynomial whose vanishing locus describes the 
Calabi--Yau manifold we are going to study. Near the conifold 
limit, it can be written as an $ALE$ fibration containing the 
Seiberg--Witten curve of $SU(2)$ SYM with a single massive 
hypermultiplet in the doublet representation. In section 3, we define 
a basis of cycles to calculate the periods of the Calabi--Yau. 
In section 4, we set up an expansion around the conifold singularity
and show that the \Ka potential of the supergravity theory reproduces 
the rigid \Ka potential when the Planck mass is taken to 
infinity. Conclusions are presented in section 5.

\section{The Calabi--Yau as a $K3$ and $ALE$ fibration}

The complex manifold we are concerned with, is described as a 
hypersurface in a weighted projective ambient space with 
homogeneous coordinates $(x_1,x_2,x_3,x_4,x_5)$ carrying 
weights $(1,2,3,3,3)$. In addition, we impose the following
global identifications:
\beq
x_j \cong \mbox{exp} ({{2 \pi i}\over{12}} n_j) x_j, \label{ident}
\eeq
with 
\bear
(n_1,n_2,n_3,n_4,n_5) &=& m_0 (1,2,3,3,3) + m_1 (2,-2,0,0,0) \nn \\ 
&& + m_2 (0,0,3,0,-3) + m_3 (0,0,3,-3,0),  
\eear
where $m_1,m_2,m_3 \in \Zbar$. $m_0$ can be complex as well. We then 
define the manifold $X^{\ast}_{12}[1,2,3,3,3]$ by the polynomial 
constraint $W=0$, where 
\bear
W &=& -{{\Lambda^6}\over{64}}B x_1^{12} + {{\Lambda^3}\over{8}}Bm x_1^8 x_2^2 
- {{1}\over{4}}\psi_s x_1^4 x_2^4 + {{1}\over{8}}B x_2^6 \nn\\
&& + {{1}\over{4}} x_3^4 + {{1}\over{4}} x_4^4 + {{1}\over{4}} x_5^4
- \psi_0 x_1 x_2 x_3 x_4 x_5. \label{CY3}
\eear
This polynomial has weight $12=1+2+3+3+3$, which ensures the 
vanishing of the first Chern class, so we are dealing with a 
Calabi--Yau threefold ($CY3$). Up to some rescalings, it is the most 
general polynomial of degree $12$ in these variables. The 
coefficients of $x_3^4$, $x_4^4$ and $x_5^4$ have already been 
fixed by rescaling the corresponding variables. We could also 
have fixed the prefactor of the first and fourth monomials, but 
it will prove useful to keep them in place for the moment. 
$\psi_s$, $\psi_0$ and $m$ are coordinates on the moduli space.

It is easy to see that our $CY3$ is a $K3$ fibration. To this end, we 
introduce the new coordinates $\zeta$ and $x_0$, the first of 
which is invariant under the identifications (\ref{ident}). It
will play the role of base space variable. The coordinate
transformation is
\bear
\zeta = {{x_2^2}\over{x_1^4}}; && x_0 = x_1 x_2. 
\eear
The polynomial (\ref{CY3}) can now be written as
\beq
W_{K3} = {{1}\over{4}} B' x_0^4 
+ {{1}\over{4}} x_3^4 + {{1}\over{4}} x_4^4 + {{1}\over{4}} x_5^4
- \psi_0 x_0 x_3 x_4 x_5,   \label{K3}
\eeq
with
\beq
B' = {{B}\over{2}} (\zeta + \Lambda^3 {{m}\over{\zeta}} 
- {{\Lambda^6}\over{8}} {{1}\over{\zeta^2}}) - \psi_s. 
\eeq

Another $K3$ fibration, but with the same fibre, namely 
$X^{\ast}_8[1,1,2,2,2]$, has been studied extensively in \cite{rigidlim}.
There it was shown that the $K3$ manifold (\ref{K3}) develops a conifold
singularity in $B'=\psi_0^4$ and a large complex structure singularity
in $B'=0$. Therefore, the $CY3$ we are studying here, generically has six 
points in base space where the $K3$ fibre becomes singular. One can easily
check that the $CY3$ itself becomes singular for $B=0$. In this case,
it contains a curve of singularities, parametrized by the base space
coordinate $\zeta$.

We can now expand the $CY3$ around this singularity. By setting 
$B=2\e$, $\psi_s+\psi_0^4=2\e u$ (thereby keeping 
$u={{\psi_s+\psi_0^4}\over{B}}$ finite) and using the same expansion for the
$K3$ coordinates as in \cite{rigidlim}, we arrive at the following local 
$ALE$ fibration \cite{klemm1, klemm2}:
\beq
W_{ALE} = \half \e \left[ \half \left(
\zeta + \Lambda^3 {{m}\over{\zeta}} 
- {{1}\over{8}} {{\Lambda^6}\over{\zeta^2}} \right) 
+ y_1^2 - u + y_2^2 + y_3^2 \right] = 0, 
\eeq
which means we are dealing with an $A_1$ singularity \cite{arnold,arnold2}. 
Apart from the terms in $y_2$ and $y_3$, we have found the Seiberg--Witten 
curve for $SU(2)$ SYM with one massive hypermultiplet in the fundamental 
representation of the gauge group:
\beq
W_{SW} = \half \left(
\zeta + \Lambda^3 {{m}\over{\zeta}} 
- {{1}\over{8}} {{\Lambda^6}\over{\zeta^2}} \right) 
+ y_1^2 - u = 0. \label{sw}
\eeq
It can be transformed into a more familiar form \cite{hanany,sw} by 
setting
\bear
y_1 = x + {{1}\over{4}}{{\Lambda^3}\over{\zeta}}; &&
\zeta = y - (x^2 - u)^2. 
\eear
We then get
\beq
y^2 = (x^2 - u)^2 - \Lambda^3 (x + m). \label{ho}
\eeq
The Seiberg--Witten meromorphic 1-form that comes with this equation is
\beq
\lambda_{SW} = {{1}\over{2\pi i}}{{x}\over{y}} 
\left( -\half{{P(x)\Lambda^3}\over{y^2-P(x)^2}} 
- 2x \right) dx, \label{hoform}
\eeq
where $P(x)=x^2-u$. The 1-form we will use in the context of the Calabi--Yau
is
\bear
\lambda &=& y_1(\zeta) {{d\zeta}\over{2\pi i \zeta}} \label{swform}\\
             &=& x(\zeta) {{d\zeta}\over{2\pi i \zeta}} 
             - {{\Lambda^3}\over{8\pi i}} d\left({{1}\over{\zeta}}\right),
\eear
It will arise as an integral over a $K3$ cycle of the Calabi--Yau holomorphic 
3-form. One has
\beq
x(\zeta) {{d\zeta}\over{2\pi i \zeta}} = \lambda_{SW} - {{1}\over{8\pi i}}
{{mdx}\over{x+m}}.
\label{extra}
\eeq

The form (\ref{hoform}) has a double pole at infinity and a first order 
pole with residue proportional to $m$ in $x=-m$, with opposite residues 
on the two sheets. As a consequence, the periods
\bear
a_D = \int_{\alpha} \lambda_{SW}, && a = \int_{\beta} \lambda_{SW},
\eear
with $\alpha$ and $\beta$ the homology 1-cycles of the hyperelliptic
surface, are not invariant under deformations of the cycles across the
pole of $\lambda_{SW}$, so $a_D$ and $a$ can make jumps proportional to
$m$ when a monodromy transformation is performed. We can just as well choose 
(\ref{swform}) as meromorphic 1-form, since the second term in (\ref{extra})
is moduli-independent and does not affect the condition that the derivative
w.r.t. the modulus should be the holomorphic form. The 1-form $\lambda$ 
has a first order pole with residue proportional to $m$ in $\zeta=0$ and 
is regular at $\zeta=\infty$ and at the branch points of the hyperelliptic 
curve.

\section{Cycles, monodromies and periods}

We begin this section with a brief review of the results of \cite{rigidlim} 
concerning the $K3$ fibre (\ref{K3}). There the periods of the holomorphic 
2-form $\hat{\Omega}^{(2,0)}$ were given in terms of solutions to the 
Picard-Fuchs equations depending on the moduli space parameter 
\beq
z = -{{B'(\zeta)}\over{\psi_0^4}}. \label{z}  
\eeq
The $K3$ large complex structure limit is at $z=0$, while the conifold 
singularities are at
$z=-1$. The non-zero periods are
\bear
\htheta_0={{1}\over{4\pi^2}}(U_1-U_2)^2, &&
\htheta_1={{i}\over{4\pi^2}}(U_1-iU_2)^2, \nn \\
\htheta_2=-{{1}\over{4\pi^2}}(U_1+U_2)^2, &&
\htheta_3=-{{i}\over{4\pi^2}}(U_1+iU_2)^2. \label{K3periods}
\eear
The functions $U_i(z)$, $i=1,2$ have the following form in the neighbourhood
of $z=\infty$:
\bear
U_1(z) & = & {{\Gamma({{1}\over{8}})\Gamma({{5}\over{8}})}
\over{\Gamma({{3}\over{4}})}} \left({{1}\over{z}}\right)^{{{1}\over{8}}}
F({{1}\over{8}},{{1}\over{8}},{{3}\over{4}};-{{1}\over{z}}), \nn \\
U_2(z) & = & {{\Gamma({{3}\over{8}})\Gamma({{7}\over{8}})}
\over{\Gamma({{5}\over{4}})}} \left({{1}\over{z}}\right)^{{{3}\over{8}}}
F({{3}\over{8}},{{3}\over{8}},{{5}\over{4}};-{{1}\over{z}}). 
\eear
In order to analytically continue them, such that they are defined one 
the whole $z$ space, one needs a cut running from $0$ to $-1$, and another 
one from $-1$ to $\infty$. The four solutions (\ref{K3periods}) satisfy 
$\sum_{k=0}^3 \htheta_k=0$, so only three of them are independent; we will 
use the first three. (A priori, one might have expected 22 periods instead 
of three, corresponding to the 22 2-cycles of a $K3$ manifold, but most of 
them are zero, namely those that correspond to the algebraic cycles.) 

A more convenient basis of periods regarding their behaviour in the 
neighbourhood of the singularities is given by
\bear
\htheta' = F \htheta, &&
F = \left( \begin{array}{ccc} -1 & 1 & 0 \\ 1 & -2 & -1 \\
1 & 0 & 0 \end{array} \right).
\eear
The intersection matrix of the associated 2-cycles is
\beq
I' = \left( \begin{array}{ccc} -2 & 0 & 1 \\ 0 & 4 & 0 \\ 1 & 0 & 0
\end{array} \right). 
\eeq
The monodromies around the singularities at $z=0$, $z=-1$, $z=\infty$
are given by
\bear
M_0' = \left( \begin{array}{ccc} 1 & 1 & -2 \\ 0 & 1 & -4 \\
0 & 0 & 1 \end{array} \right), &&
M_{-1}' = \left( \begin{array}{ccc} -1 & 0 & 0 \\ 0 & 1 & 0 \\
1 & 0 & 1 \end{array} \right), \nn \\
M_\infty' & = & \left( \begin{array}{ccc} -3 & -1 & -2 \\ 4 & 1 & 4 \\
1 & 0 & 1 \end{array} \right). \label{mon}
\eear

We now turn to the Calabi--Yau threefold $X^{\ast}_{12}[1,2,3,3,3]$, the 
$K3$ fibration that was defined in (\ref{CY3}).
We will consider the periods of the (rescaled) holomorphic 3-form,
which we define using the Griffiths map \cite{griffiths} and then split 
up into a $K3$ holomorphic 2-form and a base space 1-form:
\beq
\hat{\Omega}^{(3,0)} = {{\psi_0 |G|}\over{(2\pi i)^4}} 
\int_\Gamma {{\omega}\over{W}} = \hat{\Omega}^{(2,0)} 
{{d\zeta}\over{2\pi i \zeta}}. 
\eeq
Here $|G|$ stands for the order of the group of identifications 
(\ref{ident}); $\Gamma$ is a cycle running around the surface $W=0$ 
in the ambient space, and $\omega$ is the volume form
\beq
\omega = w_1 x^1 dx^2 \ldots dx^5 - w_2 x^2 dx^1 \ldots dx^5 + \ldots
+ w_5 x^5 dx^1 \ldots dx^4, 
\eeq
with $(w_1,\ldots,w_5)$ the weights of $(x_1,\ldots,x_5)$.

Starting from the $K3$ periods, we can define $CY3$ periods along two
kinds of cycles.\\
$\bullet$ $S^1 \times S^2$ cycles: a $K3$ cycle fibred over a closed path in 
base space running around point(s) where the $K3$ cycle shrinks to zero;\\
$\bullet$ $S^3$ cycles: a $K3$ cycle fibred over a path running between 
two points where the cycle vanishes. 

As can be seen from (\ref{z}), the branch cuts in the $z$ plane induce 
cuts in the base space of the Calabi--Yau. In our case, $z$ is given by
\beq
z = - {{{{B}\over{2}} (\zeta + \Lambda^3 {{m}\over{\zeta}} - 
{{\Lambda^6}\over{8}} {{1}\over{\zeta^2}}) - \psi_s}\over{\psi_0^4}}.
\label{zdef}
\eeq
We already mentioned that there are three points in base space where the 
$K3$ fibre develops a conifold singularity, while in three other points 
there is a large complex structure singularity. The large complex structure 
points are solutions of 
\beq
{{B}\over{2}} \left[ \zeta + \Lambda^3 {{m}\over{\zeta}} 
- {{1}\over{8}}{{\Lambda^6}\over{\zeta^2}} \right] - \psi_s = 0. 
\eeq
Let us call $e_0^1$ and $e_0^2$ the two points that go to zero and 
$e_\infty$ the point that goes to $\infty$ as $\psi_s+\psi_0^4 
\rightarrow 0$, $B \rightarrow 0$. The conifold points satisfy
\beq
{{B}\over{2}} \left[ \zeta + \Lambda^3 {{m}\over{\zeta}} 
- {{1}\over{8}}{{\Lambda^6}\over{\zeta^2}} \right] 
- (\psi_s+\psi_0^4) = 0,
\eeq
and we shall call $f^1_0$, $f^2_0$ the solutions that converge to
$e^1_0$, $e^2_0$ respectively, and $f_\infty$ the point that goes
to $e_\infty$, as $\psi_0 \rightarrow 0$.

There are six branch cuts in the $CY3$ base space. Two of them run 
from $e^{1,2}_0$ to $f^{1,2}_0$, another two connect $f^{1,2}_0$ to zero,
and finally there are cuts between $e_\infty$ and $f_\infty$ and from
$f_\infty$ to $\infty$.

We can now construct the $CY3$ periods (see fig. 1). First we define 
the periods associated to cycles of topology $S^1 \times S^2$. They are 
given by integrating $\htheta_0'$, $\htheta_1'$, and $\htheta_2'$
around a path $C$ encircling the points zero, $f^{1,2}_0$ and $e^{1,2}_0$:
\bear
T_0 & = & {{1}\over{2\pi i}} 
\int_{C} {{d\zeta}\over{\zeta}} \htheta_0', \nn\\
T_1 & = & {{1}\over{2\pi i}} 
\int_{C} {{d\zeta}\over{\zeta}} \htheta_1', \nn\\
T_2 & = & {{1}\over{2\pi i}} 
\int_{C} {{d\zeta}\over{\zeta}} \htheta_2'. 
\eear

Next, we construct the periods along the $S^3$ cycles. From the
monodromy matrices (\ref{mon}), we can read off that $\htheta'_2$
gets a contribution $\htheta'_0$ when running around the conifold 
singularity $z=-1$. From the Picard--Lefschetz formula, it follows that
the only extra cycles that can appear are vanishing cycles, so 
$\htheta'_0$ must vanish at the conifold points $f^{1,2}_0$, 
$f_\infty$. Thus, we obtain an $S^3$ cycle as follows:
\beq
V_0 = {{1}\over{2\pi i}} \int_{f^1_0}^{f_\infty} 
{{d\zeta}\over{\zeta}} \htheta_0'. 
\eeq
Now look at the monodromy at $z=0$. By the same reasoning as in the case 
of the conifold point, we find that $\htheta_1'$ and $\htheta_2'$ must 
vanish at the large complex structure points. So we define the periods
\bear
V_1^\infty &=& {{1}\over{2\pi i}} \int_{e^1_0}^{e_\infty} 
{{d\zeta}\over{\zeta}} \htheta_1', \nn \\
\tilde{V}_1^\infty &=& {{1}\over{2\pi i}} \int_{e^2_0}^{e_\infty} 
{{d\zeta}\over{\zeta}} \htheta_1', \nn \\
V_2^\infty &=& {{1}\over{2\pi i}} \int_{e^1_0}^{e_\infty} 
{{d\zeta}\over{\zeta}} \htheta_2', \nn \\
\tilde{V}_2^\infty &=& {{1}\over{2\pi i}} \int_{e^2_0}^{e_\infty} 
{{d\zeta}\over{\zeta}} \htheta_2', 
\eear
We let the paths of $V_{1,2}^\infty$ intersect that of $V_0$. 
Later on, it will be useful not to consider these periods but the sums
and differences:
\bear
V_1^+ &=& V_1^\infty + \tilde{V}_1^\infty \nn\\
V_2^+ &=& V_2^\infty + \tilde{V}_2^\infty \nn\\
V_1^- &=& V_1^\infty - \tilde{V}_1^\infty \nn\\
V_2^- &=& V_2^\infty - \tilde{V}_2^\infty 
\eear

The eight cycles associated to the periods $T_0$, $V_0$, $T_1$, $T_2$, 
$V_{1,2}^-$, $V_{1,2}^+$ together yield  an invertible intersection 
matrix for the Calabi--Yau. This is precisely the number of cycles
we need for a basis, since $h^{2,1}=3$, so that the third Betti number is
$b_3 = h^{3,0}+h^{2,1}+h^{1,2}+h^{0,3} = 8$. We can bring the 
intersection matrix in block-diagonal form by defining 
\bear
V_2^{+'} & = & 2V_2^+ + V_0 + \tilde{V}_0 - {{3}\over{2}} T_0; 
\label{V2def}\\
T_2' & = & 2T_2 + T_0, \nn \\
V_1^{+'} & = & V_1^+ + 2T_2', \\  
V_1^{-'} & = & V_1^- - \half T_1, \\ 
V_2^{-'} & = & 2V_2^- - \half T_2', \label{redef} \\
\eear
where $\tilde{V}_0$ is the integral of $\htheta'_0$ over a base space
path running from $f_0^2$ to $f_\infty$.

\begin{figure}
\begin{center}
\setlength{\unitlength}{1cm}
\begin{picture}(10,7)
\put(0,0){\epsfig{file=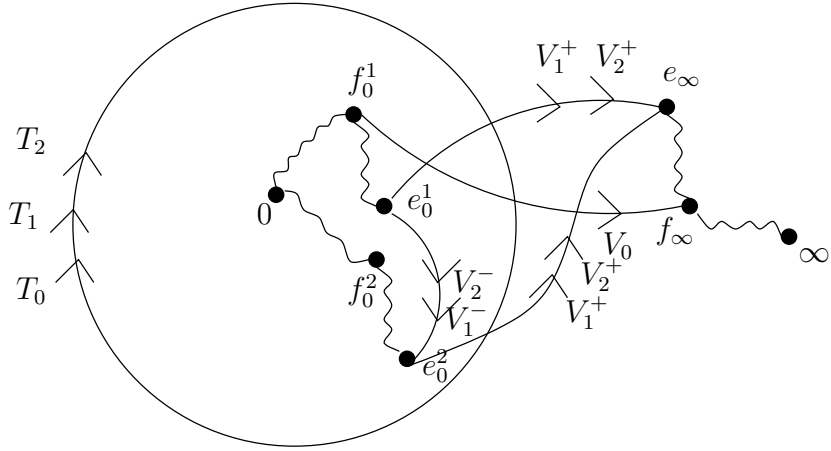,width=10cm}}
\put(7.4,2.6){$V_0$}
\put(7.3,5.1){$V_2^+$}
\put(6.5,5.1){$V_1^+$}
\put(-0.4,2){$T_0$}
\put(-0.5,3){$T_1$}
\put(-0.4,4){$T_2$}
\put(5.3,1.6){$V_1^-$}
\put(5.4,2.1){$V_2^-$}
\put(10,2.5){$\infty$}
\put(8.1,2.8){$f_\infty$}
\put(4,2){$f_0^2$}
\put(5,1){$e_0^2$}
\put(8.2,4.9){$e_\infty$}
\put(2.8,3){$0$}
\put(4,4.8){$f_0^1$}
\put(4.8,3.2){$e_0^1$}
\put(7.1,2.2){$V_2^+$}
\put(6.9,1.7){$V_1^+$}
\end{picture}
\caption{The base space paths of the $CY3$ cycles.}
\end{center}
\end{figure}
 
In the basis
\beq
\crond = \{ T_0, V_0, T_1, V_1^{+'}, T_2', V_2^{+'}, V_1^{-'}, V_2^{-'} \} 
\label{basis}
\eeq
the intersection matrix is given by 
\beq
I_{CY3} = \left( \begin{array}{cccccccc}
 0  & -2  &  0  &  0  &  0  &  0  &  0  &  0 \\ 
 2  &  0  &  0  &  0  &  0  &  0  &  0  &  0 \\
 0  &  0  &  0  &  8  &  0  &  0  &  0  &  0 \\
 0  &  0  & -8  &  0  &  0  &  0  &  0  &  0 \\
 0  &  0  &  0  &  0  &  0  & -2  &  0  &  0 \\
 0  &  0  &  0  &  0  &  2  &  0  &  0  & -1 \\
 0  &  0  &  0  &  0  &  0  &  0  &  0  & -4 \\
 0  &  0  &  0  &  0  &  0  &  1  &  4  &  0 \\
\end{array} \right). \label{intersection}
\eeq
For the cycles that do not intersect in a branch point, the intersection
is simply equal to the sum of the intersections of the base space paths, each 
multiplied by the intersection of the fibred $K3$ cycles. A little care is 
needed in calculating the intersection of, for example, $V_1^-$ with $V_1^+$. 
To do this, one can replace $V_1^-$ by an 8-shaped path running around the 
large complex structure points $e_0^1$ and $e_0^2$. The $K3$ cycle associated 
to $\htheta'_0 - \half \htheta'_1$ is transported in counter-clockwise 
direction around $e_0^1$. When crossing the cut, it is transformed into 
$\htheta'_0 + \half \htheta'_1$. This is transported around $e_0^2$ in 
clockwise direction, changing it into $\htheta'_0 - \half \htheta'_1$ 
again. $V_1^+$ intersects this cycle in two base space points, which 
together give the result $4$ for the intersection.

\section{The rigid limit}

As it is well-known, the complex scalars of vector multiplets in
$N=2$ rigid supersymmetry as well as in supergravity behave as coordinates
on a manifold with special \Ka geometry \cite{dwlsvp, dwvp}. The metric 
representing the couplings of these scalars is given in terms of a \Ka 
potential, which can be expressed as
\bear
K(u,\bar{u}) = i\langle V(u),\bar{V}(\bar{u}) \rangle; &&  
\krond(z,\bar{z}) = -\mbox{log} (-i\langle v(z),\bar{v}(\bar{z}) \rangle) 
\eear
for the rigid and local geometry respectively. The period vector
$V(u)$ (resp. $v(z)$) is a holomorphic function of $r$ (resp. $n$) 
complex scalars $\{ u^i \}$ (resp. $\{ z^\alpha \}$).
It has $2r$ (resp. $2(n+1)$) components. The symplectic inner product
$\langle . , . \rangle$ is defined by
\bear
\langle V,W \rangle = V^T Q^{-1} W; &&
\langle v,w \rangle = v^T q^{-1} w, 
\eear
where $Q$ and $q$ are real, invertible, antisymmetric matrices. $q$
will be the Calabi--Yau intersection matrix.

To find the rigid limit of the local geometry, one divides the local
coordinates $\{z^\alpha\}$ in a set $\{u^i\}$, which will become the rigid 
coordinates, and a parameter $\e$ such that $\e \rightarrow 0$
in the conifold limit. Assume that the period vector can be decomposed as 
\beq
v = v_0(\e) + \e^a v_1(u) + v_2(\e,u), \label{perexp}
\eeq
where $v_0$ contains the dominant and constant pieces and is independent of 
the surviving moduli, while $v_1$ does not depend on $\e$ and is such that 
the derivatives w.r.t. the moduli form a matrix of rank $r$, the number of 
$\{u^i\}$. For the examples of \cite{rigidlim}, it was found that
\beq
\langle v,\bar{v} \rangle = iM^2(\e) + f(\e,u) - \bar{f}(\bar{\e},\bar{u})
+ |\e|^{2a} \langle v_1(u),\bar{v}_1(\bar{u}) \rangle
+ R(\e,\bar{\e},u,\bar{u}), 
\eeq
where the function $f$ is holomorphic in $u$ and
\beq
iM^2(\e) = \langle v_0(\e),\bar{v}_0(\bar{\e}) \rangle,
\eeq
with
\bear
{{|\e|^{2a}}\over{M^2}} \rightarrow 0; &&
{{R}\over{|\e|^{2a}}} \rightarrow 0
\eear
as $\e \rightarrow 0$. $a$ is some real number which could be normalized to 
one. Assuming this structure is realized, one has 
\bear
\krond & = & - \mbox{log} (M^2) + {{i}\over{M^2}} F(\e,u) 
- {{i}\over{M^2}} \bar{F}(\bar{\e},\bar{u})
+ i{{|\e|^{2a}}\over{M^2}} \langle v_1(u),\bar{v}_1(\bar{u}) \rangle 
+ \ldots \nn \\
       & = & - \mbox{log} (M^2)  + {{i}\over{M^2}} F(\e,u) 
- {{i}\over{M^2}} \bar{F}(\bar{\e},\bar{u})
+ {{|\e|^{2a}}\over{M^2}} K(u,\bar{u}), 
+ \ldots \label{localrigid}
\eear
with $K(u,\bar{u})$ the rigid \Ka potential. The $F$ terms amount to 
a \Ka transformation, provided the $u$-dependent parts of the function
$f$ are of higher order than $|\e^a|$. 

Let us consider the period vector. We again use the results of 
\cite{rigidlim}, where the exact expressions for the $K3$ periods
can be found. Expanding in powers of $1+z \sim \e$, one has
\bear
\htheta_0' & = & \eta \sqrt{1+z} + \orond(\e^{{3}\over{2}}), \nn\\
\htheta_1' & = & k_1 + \half l_1 (1+z) 
+ \orond(\e^2), \label{k3exp} \nn \\
2\htheta_2' + \htheta_0' & = & k_2 + \half l_2 (1+z)
+ \orond(\e^2), 
\eear
where $k_1$ and $k_2$ are constants, and the parameter $z$ was
defined in (\ref{zdef}). In the rigid limit, $z$ has the expansion 
\beq
z = -1 + 2\te (u - \xi) + \orond(\tilde{\e}^2), \label{zexp}
\eeq
where we have defined
\bear
\te = -{{\e}\over{\psi_s}}; &&
\xi = \half\left(\zeta + \Lambda^3 {{m}\over{\zeta}} 
- {{\Lambda^6}\over{8}}{{1}\over{\zeta^2}} \right). \label{xidef}
\eear

Instead of directly evaluating integrals over the $K3$ periods, it is 
useful to first get a view on the general structure of the period vector 
in terms of $\te$, by calculating the $\te$ monodromies of 
the $CY3$ periods. From the way the periods transform under $\te 
\rightarrow e^{2\pi i} \te$, one can deduce what the expansion in 
$\te$ looks like \cite{arnold}. 

When we turn $\te$ in the complex plane by $\te \rightarrow 
e^{i \theta} \te$, where $\theta$ runs from 0 to $2\pi$, the conifold 
points $f_0^{1,2}$ and $f_\infty$ remain fixed. $e_0^{1,2}$ and $e_\infty$
to lowest order in $\te$ are given by
\beq
e_0^1 = -i{{3}\over{2}}{{\Lambda^3}\over{\sqrt{2}}} \, \te^\half; \,\,\,\,\,\,
e_0^2 = i{{3}\over{2}}{{\Lambda^3}\over{\sqrt{2}}}\, \te^\half; \,\,\,\,\,\,
e_\infty = -{{1}\over{\te}}. \label{lcsexp}
\eeq
This means that under a full rotation of $\te$, $e_0^1$ and $e_0^2$ will get
interchanged by turning around $0$ in counter-clockwise direction, 
while $e_\infty$ makes a full circle around $\infty$. When 
calculating the transformation of a $CY3$ cycle, one also has to take into 
account the $K3$ cycle fibres. A $K3$ cycle $c$ gets transformed into 
$M_{-1}' c$. 

In the basis (\ref{basis}), the $\te$ monodromy is given by
\beq
v \rightarrow M_{CY3}^{\te} v, 
\eeq
with
\beq
M_{CY3}^{\te} = \left( \begin{array}{cccccccc}
-1  &  0  &  0  &  0  &  0  &  0  &  0  &  0 \\
 0  & -1  &  0  &  0  &  0  &  0  &  0  &  0 \\
 0  &  0  &  1  &  0  &  0  &  0  &  0  &  0 \\
 0  &  0  &  3  &  1  &  0  &  0  &  0  &  0 \\
 0  &  0  &  0  &  0  &  1  &  0  &  0  &  0 \\
 0  &  0  &  0  &  0  &  3  &  1  &  0  &  0 \\
 0  &  0  &  0  &  0  &  0  &  0  & -1  &  0 \\
 0  &  0  &  0  &  0  &  0  &  0  & -1  & -1 
\end{array} \right). \nn
\eeq
A slight subtlety about $V_2^{-'}$ is that it transforms as
\beq
V_2^{-'} \rightarrow -V_2^{-'} + W_0,
\eeq
where $W_0$ is the integral of $\htheta'_0$ over two paths that run from
$f_0^1$ to $f_0^2$ on opposite sides of the cuts. By checking the 
intersections of the associated $CY3$ cycle with the basis of $CY3$ cycles 
or by directly deforming it, one finds that it is equal to $-V_1^{-'}$.

The above expressions for the $\te$ monodromy are almost sufficient 
to derive the structure of the period vector in terms of $\te$. It 
looks like
\beq
\left( \begin{array}{c}
T_0 \\ V_0 \\ T_1 \\ V_1^{+'} \\ T'_2 \\ V_2^{+'} \\ V_1^{-'} \\ V_2^{-'} 
\end{array} \right)
=
\left( \begin{array}{c}
\te^{\half} (V_1(u) + \orond (\te)) \\
\te^{\half} (V_2(u) + \orond (\te)) \\
-k_1 - \te l_1 u + \orond (\te^2) \\
-{{3}\over{2\pi i}}\mbox{log}\te\,(k_1 + \te u l_1 + \orond(\te^2))
+ k'_1 + l'_1 u \te + \orond(\te^2) \\
-k_2 - \te l_2 u + \orond (\te^2) \\
-{{3}\over{2\pi i}}\mbox{log} \te \, (k_2 + l_2 u \te + \orond(\te^2))
+ k'_2 + l'_2 u \te + \orond(\te^2) \\ 
n \te^\half + \orond(\te^{{3}\over{2}}) \\
{{1}\over{2 \pi i}} \mbox{log}\te \, (n \te^\half 
+ \orond(\te^{{3}\over{2}})) + V(u)\te^\half + \orond(\te^{{3}\over{2}})
\end{array} \right). \label{periodvector}
\eeq
The constants $k_{1,2}$ and $l_{1,2}$ are the same as the ones in 
(\ref{k3exp}). The expressions for $T_1$ and $T'_2$ can easily be
found by direct integration.

The periods $V_1^{-'}, V_2^{-'}, V_1^{+'}, V_2^{+'}$ need some explanation. 
From the $\te$ monodromy of $V_1^{-'}$, one finds that it must have the 
structure
\beq
V_1^{-'} = \te^\half n(u) + \orond(\te^{{3}\over{2}}).
\eeq
We can also evaluate it directly. Writing $\htheta'_1(z(\zeta))=f(\zeta)$, 
we have, using (\ref{lcsexp}), (\ref{k3exp}) and (\ref{zexp}),
\bear
\int_{e_0^1}^{e_0^2} {{d\zeta}\over{2\pi i \zeta}} f(\zeta) - \half T_1 
&=& -\half (f(0) + f'(0) 2 \te u + \orond(\te^2)) 
+ \half (k_1 + l_1 \te u + \orond (\te^2)) \nn\\
&& + \sum_{i=1}^\infty \sum_{j=1}^i \sum_{k=0}^j \sum_{l=0}^k A_{ijkl}
u^{i-j} [(\te^\half)^{2i-2j+k+2l} - (-\te^\half)^{2i-2j+k+2l}] \nn\\
&=& n \te^\half + \orond(\te^{{3}\over{2}}), 
\eear
where $n$ is indeed a $u$-independent constant. The minus sign in the
first line arises because we are integrating on the `lower' side of
zero.

$V_2^{-'}$ is an integral of $\half (2\htheta'_2+\htheta'_0)$ over a path
from $e_0^1$ to $e_0^2$ running around the outer side of $f_0^1$, plus 
$\half (2\htheta'_2+\htheta'_0)-\htheta'_0$ integrated over a path between
the same points around the outer side of $f_0^2$ (for the definition
of $V_2^{-'}$, see (\ref{redef})). The integrals of 
$\half(2\htheta'_2+\htheta'_0)$ add up to something of the form 
$\tilde{n} \te^\half + \orond(\te^{{3}\over{2}})$, where $\tilde{n}$ is a 
constant. The integral of $-\htheta'_0$ over a path from $e_0^1$ to $e_0^2$ 
around $f_0^2$ can be expected to depend on $u$ in a non-trivial way, and it
transforms to minus itself plus $W_0$ under an $\te$ monodromy. As we 
mentioned before, $W_0$ is nothing but $-V_1^{-'}$. Putting everything 
together, we find that $V_2^{-'}$ is of the form
\beq
V_2^{-'} = {{1}\over{2\pi i}} \log \te \, (n \te^\half 
+ \orond(\te^{{3}\over{2}})) + V(u)\te^\half + \orond(\te^{{3}\over{2}}).
\eeq

By direct integration, it is easy to see that the constants $k'_1$ 
and $l'_1$ in the expansion of $V_1^{+'}$ are $u$-independent. To check 
this for the constants $k'_2$ and $l'_2$ in $V_2^{+'}$, one writes
\beq
V_2^{+'} = \left(\int_{e_0^1}^{e_\infty} + \int_{e_0^2}^{e_\infty} \right)
(2\htheta'_2 + \htheta'_0) 
+ \left(-\int_{e_0^1}^{e_\infty} - \int_{e_0^2}^{e_\infty} 
+ \int_{f_0^1}^{f_\infty} + \int_{f_0^2}^{f_\infty} 
- {{3}\over{2}} \int_C \right) \htheta'_0. 
\eeq
(See eq. (\ref{V2def}) for the definition of $V_2^{+'}$.) Only the first
two integrals need to be considered; the rest are associated to cycles that,
after some deformations, can be seen to cancel against each other in the 
rigid limit (see fig. 2).

\begin{figure}
\begin{center}
\setlength{\unitlength}{1cm}
\begin{picture}(10,7)
\put(0,0){\epsfig{file=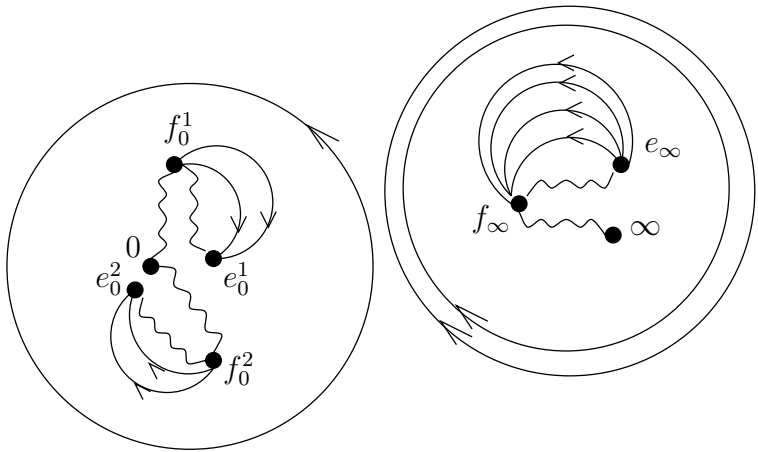,width=10cm}}
\put(1.6,2.6){$0$}
\put(1.2,2.2){$e_0^2$}
\put(2.9,1){$f_0^2$}
\put(2.1,4.2){$f_0^1$}
\put(2.9,2.2){$e_0^1$}
\put(8.3,2.9){$\infty$}
\put(8.5,4){$e_\infty$}
\put(6.2,3){$f_\infty$}
\end{picture}
\caption{The base space paths of the cycles that cancel in the rigid 
limit in the calculation of $V_2^{+'}$.}
\end{center}
\end{figure}

Now that we have determined the $\te$ expansions of the periods, we 
can bring the intersection matrix of the $CY3$ into a particularly simple 
form by performing another basis transformation (which would, however, 
have spoiled the Jordan form of the $\te$ monodromy matrix):
\beq
V_2^{+''} = V_2^{+'} - {{1}\over{4}} V_1^{-'}.
\eeq
In the basis 
\beq
\crond' = \{ T_0, V_0, T_1, V_1^{+'}, T_2', V_2^{+''}, V_1^{-'}, V_2^{-'} \} 
\eeq
we now get the following expression for the inverse intersection matrix:
\beq
q^{-1} = {{1}\over{8}} \left( \begin{array}{cccccccc}
 0  &  4  &  0  &  0  &  0  &  0  &  0  &  0  \\
-4  &  0  &  0  &  0  &  0  &  0  &  0  &  0  \\
 0  &  0  &  0  & -1  &  0  &  0  &  0  &  0  \\
 0  &  0  &  1  &  0  &  0  &  0  &  0  &  0  \\
 0  &  0  &  0  &  0  &  0  & -4  &  0  &  0  \\
 0  &  0  &  0  &  0  &  4  &  0  &  0  &  0  \\
 0  &  0  &  0  &  0  &  0  &  0  &  0  &  2  \\
 0  &  0  &  0  &  0  &  0  &  0  & -2  &  0 
\end{array} \right).  
\eeq
From (\ref{periodvector}), we find
\bear
v_0 &=& \left( \begin{array}{c}
0 \\ 0 \\ -k_1 \\ -{{3}\over{2\pi i}}\mbox{log}\te\, k_1 + k'_1 \\
-k_2 \\ -{{3}\over{2\pi i}}\mbox{log} \te \, k_2 + k''_2  
\\ 0 \\ 0
\end{array} \right); \,\,\,\,
v_1 = \left( \begin{array}{c}
V_1(u) \\ V_2(u) \\ 0 \\ 0 \\ 0 \\ -\half n \\ n \\ 
{{1}\over{2 \pi i}} \mbox{log}\te \, n + V(u) 
\end{array} \right); \nn\\
v_2 &=& \left( \begin{array}{c}  
\orond (\te^{{3}\over{2}}) \\ \orond (\te^{{3}\over{2}}) \\
- \te l_1 u + \orond (\te^2) \\
-{{3}\over{2\pi i}}\mbox{log}\te\,(\te u l_1 + \orond(\te^2))
+ l'_1 u \te + \orond(\te^2) \\
- \te l_2 u + \orond (\te^2) \\
-{{3}\over{2\pi i}}\mbox{log} \te \, (l_2 u \te + \orond(\te^2))
+ l''_2 u \te + \orond(\te^2) \\ 
\orond(\te^{{3}\over{2}}) \\
\orond(\te^{{3}\over{2}})
\end{array} \right). 
\eear
It will be clear that $i \langle v_0,\bar{v}_1 \rangle = 0$, up to a 
constant, which will lead to a constant \Ka transformation. On the other 
hand, 
\beq
\langle v_{1},\bar{v}_{1} \rangle =  
\eta^2 \left(
\int_C \lambda 
\int_{f_0^1}^{f_\infty} \bar{\lambda}
- \int_{f_0^1}^{f_\infty} \lambda 
  \int_C \bar{\lambda} \right),  \label{rigidkahlerpot} 
\eeq
up to a $u$-dependent expression, which will again lead to a \Ka 
transformation. (\ref{rigidkahlerpot}) is equal to the rigid \Ka potential 
$K(u,\bar{u})$, up to a prefactor which can be absorbed into the holomorphic 
3-form of the $CY3$ by a rescaling. As we already noted in section 2, the form
$\lambda$ can be used as Seiberg-Witten meromorphic 1-form \cite{marshakov}.
It is the analogue of what one gets from the $CY3$ holomorpic 3-form in
the case of gauge theories without matter \cite{klemm1, klemm2}. 

Because of the $l$ terms in $v_2$, the 
products $\langle v_0,\bar{v}_2 \rangle$ and $\langle v_1,\bar{v}_2 \rangle$ 
also give a contribution that amounts to a \Ka transformation. 
The product $\langle v_2,\bar{v}_2 \rangle$ gives non-holomorphic 
contributions in $u$ that are, however, of higher order than $\te$.
Thus, we indeed find the structure
\beq
\krond = - \mbox{log} (M^2) 
+ {{|\e|}\over{M^2}} K(u,\bar{u}) 
+ {{i}\over{M^2}}F(\te,u) -{{i}\over{M^2}}\bar{F}(\bar{\te},\bar{u}) + 
\orond(|\te^2|). 
\eeq

\section{Conclusions}

We considered a type IIB compactification on a $K3$ fibration which
near the conifold singularity contained the SW curve for $SU(2)$ SYM with 
a massive quark hypermultiplet. We performed the rigid limit and explicitly
showed how the field theory degrees of freedom decouple from the 
gravitational ones. 

The $K3$ fibre was already encountered in \cite{rigidlim}, 
where a $CY3$ for pure $SU(2)$ SYM was studied. We introduced matter
by changing the way the $K3$ was fibred over base space. We then showed
how to treat the $K3$ fibration globally, instead of resorting to
an $ALE$ approximation. This allowed for an explicit embedding into a 
supergravity model. The structure that had been found for the local 
\Ka potential of pure gauge supergravity theories in terms of the rigid 
\Ka potential, was not spoiled by the introduction of matter.

Our results could easily be extended to introduce matter in any $K3$ 
fibration. It should not be difficult to repeat the procedure in the case of 
a larger gauge group and more matter. Thus, knowing the periods 
and monodromies of the $K3$ fibre should allow to quickly obtain information
about a variety of models.

\medskip
\section*{Acknowledgements}

I would like to thank M. Bill\'o, F. Denef and A. Van Proeyen for very
helpful discussions.

\end{document}